\documentclass[article]{JHEP3}

\usepackage{graphicx}
\usepackage{amsmath,epsfig}
\usepackage{amssymb,amsfonts}
\usepackage{latexsym}

\relax

\def\be{\begin{equation}}
\def\ee{\end{equation}}

\newcommand{\la}{\lambda}

\newcommand{\bear}{\begin{eqnarray}}
\newcommand{\bea}{\begin{eqnarray}}
\newcommand{\eear}{\end{eqnarray}}
\newcommand{\eea}{\end{eqnarray}}

\def\hri#1#2{\href{http://arxiv.org/abs/#1}{[ArXiv:#1]#2}}
\def\hre#1#2{\href{http://arxiv.org/abs/#1/#2}{[ArXiv:#1/#2]}}

\newbox\pippobox

\def\II{\relax{\rm I\kern-.18em I}}

\def\l{\lambda}
\def\m{\mu}
\def\n{\nu}

\def\sp{\;\;\;,\;\;\;}

\def\h{\kappa}
\def\gf{w}

\title{V-QCD: Spectra, the dilaton and the S-parameter}

\author{Daniel Are\'an$^a$, Ioannis Iatrakis$^c$, Matti J\"arvinen$^c$ and Elias Kiritsis$^{b,c}$\\
~\\
$^a$ International Centre for Theoretical Physics (ICTP)
and INFN - Sezione di Trieste \\
Strada Costiera 11; I 34014 Trieste, Italy \\
~\\
$^b$ \href{http://www.apc.univ-paris7.fr}{APC, Universit\'e Paris 7, Diderot},\\
 CNRS/IN2P3, CEA/IRFU, Obs. de Paris, Sorbonne Paris Cit\'e, \\
  B\^atiment Condorcet, F-75205, Paris Cedex 13, France (UMR du CNRS 7164)\\
~\\
$^c$ \href{http://hep.physics.uoc.gr}{Crete Center for Theoretical Physics},
Department of Physics, University of Crete, 71003 Heraklion, Greece\\
}

\preprint{CCTP-2012-19}

\abstract{Zero temperature spectra of mesons and glueballs are analyzed in
a class of holographic bottom-up models for QCD (named V-QCD),
as a function of $x={N_f\over N_c}$ with the full back-reaction included.
It is found that spectra are discrete and gapped (modulo the pions) in the QCD regime,
for $x$ below the critical value $x_c$ where the conformal transition takes place.
The masses uniformly converge to zero in the walking region $x\to x_c$
due to Miransky scaling. The ratio of masses all asymptote to non-zero constants as $x\to x_c$ and therefore there is no ``dilaton" in the spectrum.
The S-parameter is computed and found to be of $\mathcal{O}(1)$ in the walking regime.
}

\begin{document}

\maketitle 

\section{Introduction}

The dynamics of ``walking" (or nearly conformal) quantum field theories, have been the subject of intensive
study, since they have been argued to be an important ingredient, \cite{walk1,walk2,walk3} in providing viable non-perturbative mechanisms for electroweak symmetry breaking
like technicolor,  \cite{tech}.
This regime is expected to appear in standard QCD with $N_c$ colors and $N_f$ flavors, just below the boundary of the conformal window, $N_f\simeq 4N_c$, as well as in other quantum field theories, \cite{neil}.

The transition between the conformal window and QCD-like IR behavior, has been called a conformal transition \cite{conf}.
The ``walking" regime has been conjectured to display Miransky scaling, \cite{miransky}. It was recently suggested that
in holographic theories this conformal transition is associated with a violation of the BF bound in the dual bulk theory, \cite{son}.
Moreover, in QCD, this correlates with fermion bilinear operators reaching a scaling dimension equal to 2, another prerequisite of viable extended technicolor.

Apart from Miransky scaling several other phenomena have been associated with the ``walking regime" of QFT:
\begin{itemize}
 \item[(a)] The appearance of a light scalar state, the ``dilaton", due to the almost unbroken scale invariance, \cite{walk2}.
 \item[(b)] The strong suppression of the S-parameter, a crucial ingredient for the experimental viability of technicolor theories, \cite{as}.
\end{itemize}
Both issues are controversial, especially as ``walking regimes" appear at strong coupling, and perturbative techniques do not apply.

In recently studied holographic models with walking behavior, the lightest state is often a scalar, \cite{dilaton}. 
Whether this state can be identified as the dilaton is, however, a difficult question and appears to depend on the model.
The S-parameter has been studied in popular holographic bottom-up \cite{hong} as well as brane-antibrane models \cite{cobi}
with a variety of answers found.
Recently, it was argued that in a class of holographic models
the S-parameter is substantial and definitely bounded below, \cite{rubakov}.

What we plan to do in this letter is to report on these and related issues
 in a class of holographic theories  that have been proposed  recently, under the name of V-QCD, \cite{jk} which has physics that is very close to
 QCD in the Veneziano limit.

\section{V-QCD}

The class of models in question combine two sectors whose dynamics are
 inspired by string holographic models. The first is improved holographic QCD (IHQCD), which is a
  holographic model for large-N Yang Mills in 4 dimensions, \cite{ihqcd}.
  The second is a model for flavor inspired by tachyon condensation in string theory, \cite{ckp}.
The relevant fields in the gravity description that are kept in these models
(in order to describe the vacuum structure) are as follows: apart for the
five-dimensional metric, there is a scalar, (the dilaton, $\phi$) that is dual
to the YM  't Hooft coupling constant, and a complex $N_f\times N_f$ matrix field,
(the tachyon, $T_{ij}$) transforming in the $(N_f,\bar N_f)$ of the
 $U(N_f)\times U(N_f)$ flavor group. We will be working in the Veneziano limit, $N_c,N_f\to\infty$ with ${N_f\over N_c}=x$ fixed, \cite{veneziano}.

The complete action for V-QCD models can be written as
\be
 S = S_g + S_f + S_a
\ee
where $S_g$, $S_f$, and $S_a$ are the actions for the glue, flavor and CP-odd sectors, respectively\footnote{We will not discuss $S_a$ here. We will address its physics (that contains the $U(1)_A$ anomaly) in a future publication, \cite{to}.}.
The glue action was introduced in \cite{ihqcd,ihqcd2,data},
\be
S_g= M^3 N_c^2 \int d^5x \ \sqrt{-g}\left(R-{4\over3}{
(\partial\lambda)^2\over\lambda^2}+V_g(\lambda)\right) \ ,
\ee
with $\la=e^{\phi}$. The dilaton potential $V_g$ asymptotes to a constant near $\la=0$,
and diverges as $V_g\sim \l^{4\over 3}\sqrt{\log\la}$ as $\l\to \infty$, generating confinement, a mass gap, discrete spectrum and asymptotically linear glueball trajectories.

\begin{figure}[!tb]
\centering
\includegraphics[width=0.45\textwidth]{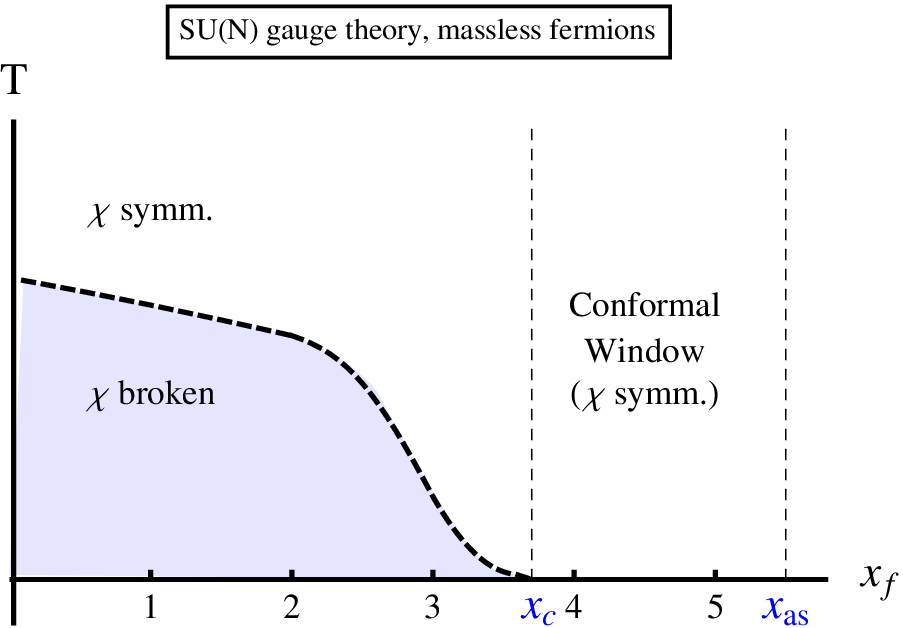}
\caption{\small Qualitative behavior of the transition temperature between the low and high
temperature phases of V-QCD matter, \cite{alho}.
 }
\label{f1}
\end{figure}

The flavor action is\footnote{This is inspired by the Sen's action for tachyon condensation, \cite{senreview}.
The non-abelian action is not well understood, but its knowledge is not needed for this paper.}
\be
S_f= - \frac{1}{2} M^3 N_c \int d^4x\, dr\,  \mathrm{Tr}\Big[V_f(\l,TT^{\dagger})
\left(\sqrt{-\det {\bf A}_L}+\sqrt{-\det {\bf A}_R}\right)\Big]\,.
\label{generalact}
\ee
The quantities inside the square roots are defined as
\be
{\bf A}_{(i)MN}=g_{MN} + \gf(\l) F^{(i)}_{MN}
+ {\kappa(\l) \over 2 } \left((D_M T)^* (D_N T)+
(D_N T)^* (D_M T)\right)
\label{Senaction}
\ee
with $(i)=L,R$, and the fields  $A_{(i)}$ as well as $T$ are $N_f \times N_f$ matrices in the flavor space.
The covariant derivative of the tachyon field is defined as
\be
D_M T = \partial_M T + i  T A_M^L- i A_M^R T \ .
\ee
The class of tachyon potentials that we will explore are
\be
V_f(\l,T)=V_{f0}(\l) e^{- \frac{a(\l)}{N_f}\mathrm{ Tr}\{T T^\dagger\}} \ .
\label{tachpot}
\ee
For the vacuum solutions (with flavor independent quark mass) we set  $T =\tau(r) \mathbf{1}_{N_f}$ where $\tau(r)$ is real, and the
flavor gauge fields are trivial.
$V_g(\l)$ has been fixed already from glue dynamics \cite{data}.
The other undetermined functions in the flavor action ($V_{f0}(\l)$, $\kappa(\l)$, $a(\l)$, $w(\la)$) must satisfy the following generic requirements:
\begin{itemize}
\item[(a)] There should be two extrema in the potential for  $\tau$: an unstable maximum at $\tau=0$ with chiral symmetry intact and a minimum at $\tau=\infty$ with chiral symmetry broken.
\item[(b)] The dilaton potential at $\tau=0$, namely $V_\mathrm{eff}(\la)=V_g(\la)-xV_{f0}(\la)$, must have a non-trivial IR extremum at $\la=\la_*(x)$ that moves from $\la_*=0$
at $x={11\over 2}$ to large values as $x$ is lowered.
\end{itemize}
In \cite{jk,alho}, several classes of potentials have been explored. The models were classified according to the IR behavior of the tachyon (I or II)
and the constant $W_0$, that controls the flavor dependence of the UV AdS scale. We refer to \cite{alho} for a detailed exposition.

At zero temperature the phase diagram as a function of $0<x<{11\over 2}$, is essentially universal. In the
region $0<x<x_c$, the (massless) theory has chiral symmetry breaking, and flows to a massless $SU(N_f)$ pion
theory in the IR. For $x_c<x<{11\over 2}$ (the conformal window) the theory flows to a non-trivial IR fixed
point and there is no chiral symmetry breaking. It was found that $x_c\simeq 4$, its precise value depends on
the details of the potential.
In the regime just below the conformal window, $x\simeq x_c$ the theory exhibits Miransky scaling,
\cite{miransky} where in particular for the chiral condensate
$\sigma\sim \Lambda_{\rm UV}^3 \exp (-{\kappa\over \sqrt{x_c-x}})$ with $\kappa$ calculable from the flavor action,
\cite{jk}.
This regime is also known as the ``walking" regime as the coupling flows to almost $\la_*$, stays there for many decades in the RG time, and then at the end the non-trivial tachyon drives the theory away from the non-trivial fixed point and towards $\la=\infty$.

At finite temperature a rich structure of hairy black holes  was found, with one and two scalar hairs, \cite{alho}.
The general structure of their phase diagram is depicted in figure \ref{f1}. The chiral restoration transition is first order at low values of $x$
but
typically becomes second order as we approach $x_c$. 
In this region extra chirally broken phases can appear
 and several extra first-order  transitions (up to two), depending on the detailed potentials.
 The transition temperatures also exhibit Miransky scaling, \cite{alho}.

\begin{figure}[t]
\begin{center}
\includegraphics[width=0.49\textwidth]{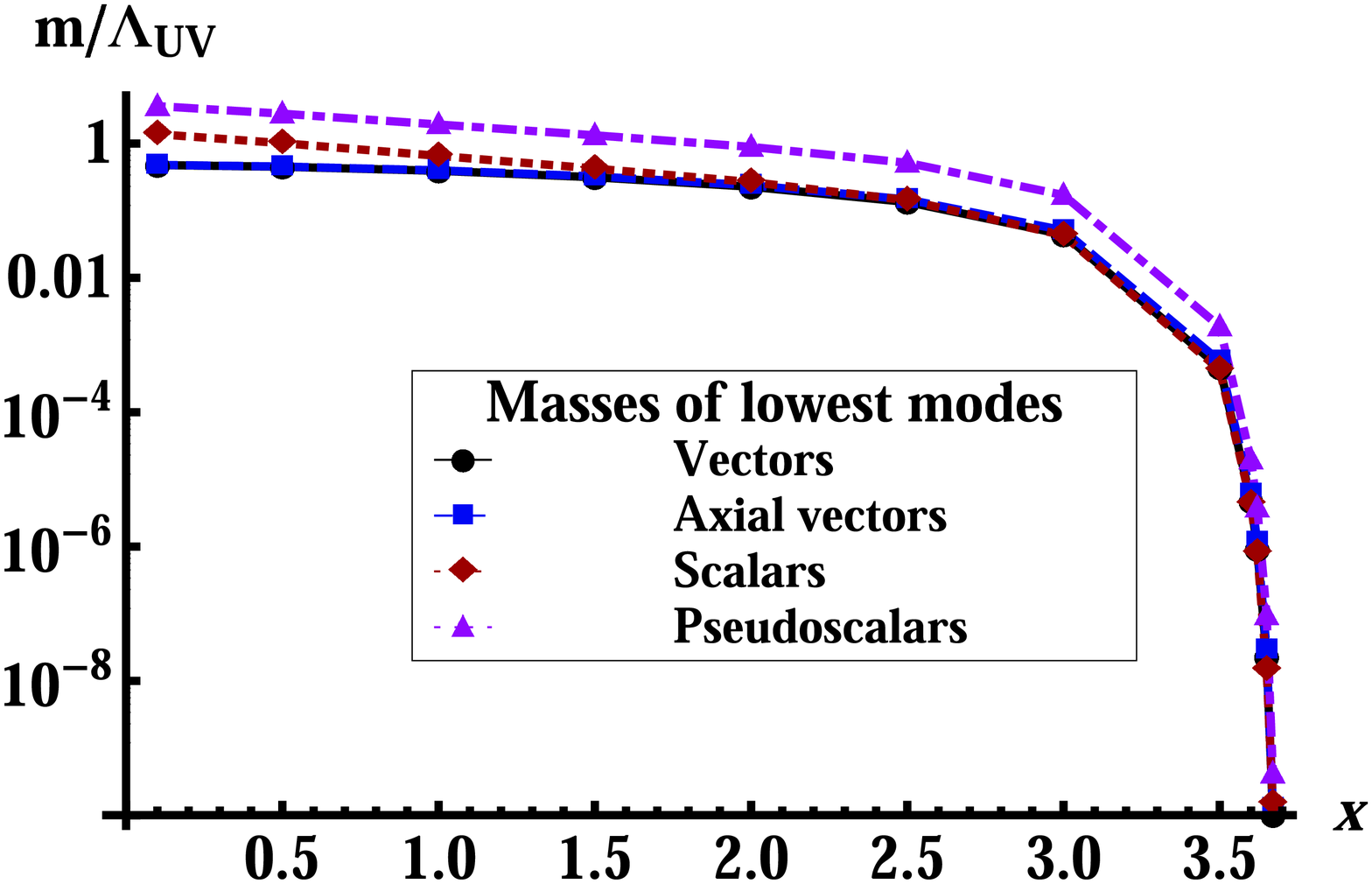}\hfill
\includegraphics[width=0.49\textwidth]{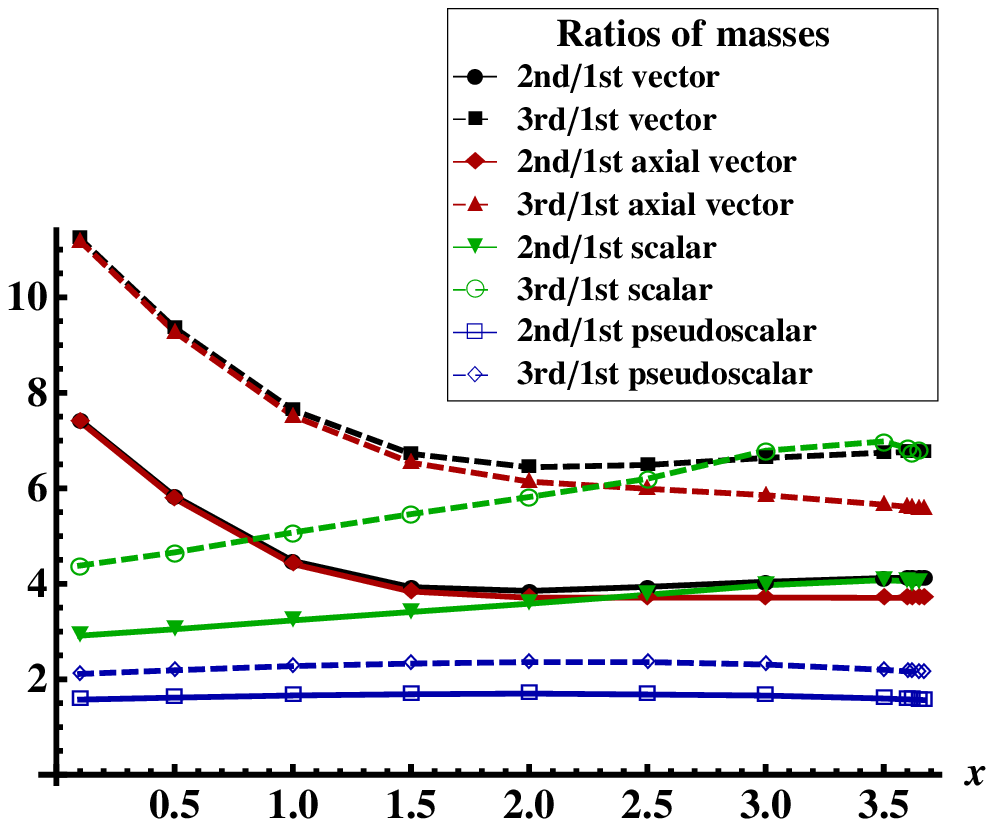}
\end{center}
\caption{\small Non-singlet meson spectra in the potential II class with Stefan-Boltzmann (SB) normalization for $W_0$ (see \cite{alho}), with $x_c \simeq 3.7001$. Left: the lowest non-zero masses of all four towers of mesons, as a function of $x$, in units of
$\Lambda_{\rm UV}$, below the conformal window. Right, the ratios of masses of up to the fourth massive states in the same theory as a function of $x$.
}
\label{f2}\end{figure}

\begin{figure}[t]
\begin{center}
\includegraphics[width=0.49\textwidth]{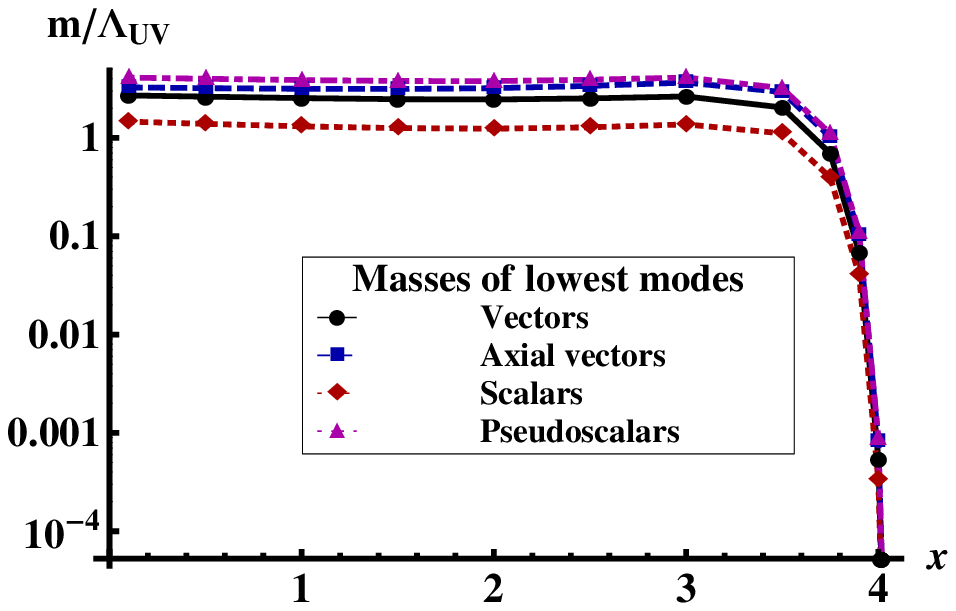}\hfill
\includegraphics[width=0.49\textwidth]{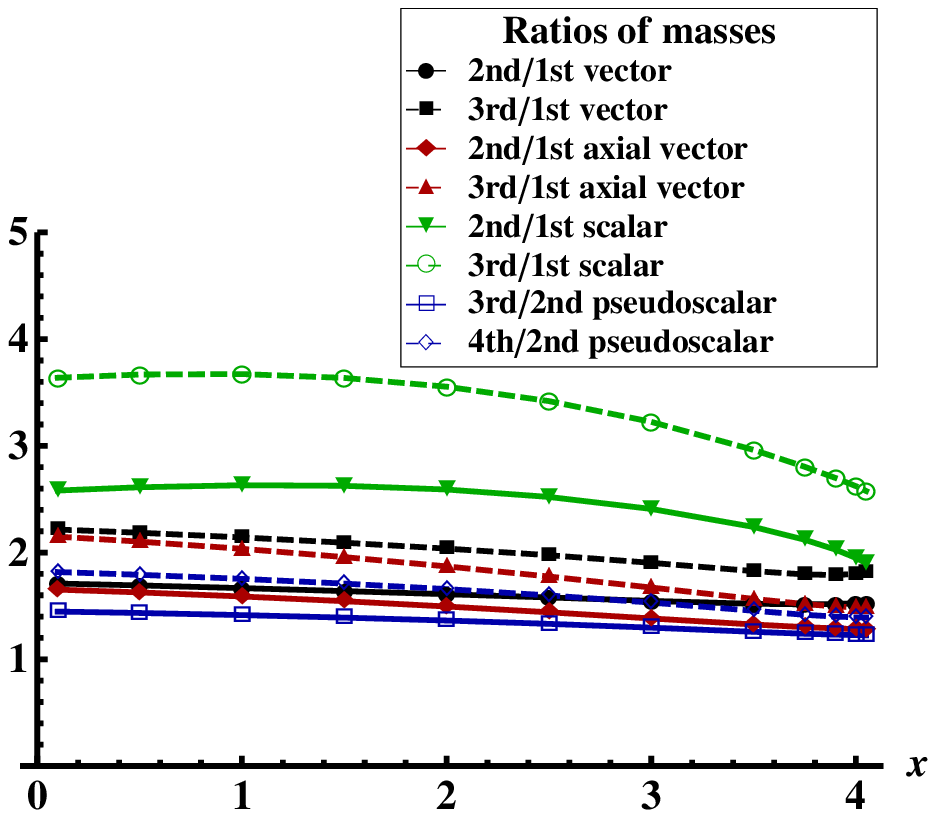}
\end{center}
\caption{\small Non-singlet meson spectra in the potential I class ($W_0={3\over 11}$), with $x_c \simeq 4.0830$. Left: the lowest non-zero masses of all four towers of mesons, as a function of $x$,
in units of $\Lambda_{\rm UV}$, below the conformal window. Right, the ratios of masses of up to the fourth massive states in the same theory as a function of $x$.
}
\label{f3}\end{figure}

\section{Quadratic fluctuations and spectra}

The purpose of this letter is to study further interesting aspects of the physics of V-QCD models and address several interesting questions
that are now accessible to calculation. In this paper we consider the theory with massless quarks.

The strategy is to consider small (quadratic) fluctuations of all the fields in V-QCD ($g_{\m\n}, \phi, T, A^{L,R}_{\m}$) around the vacuum (zero temperature) solutions of \cite{jk}, involving a Poin\-car\'e invariant metric, no vectors and radially depended scalars,
\be
ds^2=e^{2 A(r)} (dx_{1,3}^2+dr^2)\sp \phi(r)\sp T=\tau(r)~\mathbf{1}_{N_f}\;.
\label{bame}
\ee
There are several fluctuations that we will classify into two distinct classes: singlet fluctuations under the flavor group, and non-singlet fluctuations.
\begin{itemize}

\item The non-singlet fluctuations include the L and R vector meson fluctuations, packaged into an axial and
vector basis, $V_{\mu},A_{\m}$, the pseudoscalar mesons (including the massless pions), and the scalar mesons.
Their second order equations are relatively simple, and we present those of the vectors below. We work in
axial gauge $V_r=0$ and write the factorized Ansatz $V_\mu (x^\mu, r) =  \psi^V(r)\, {\cal V}_\mu (x^\mu)$.
The radial wave functions satisfy

 \begin{figure}[!tb]
\begin{center}
\includegraphics[width=0.49\textwidth]{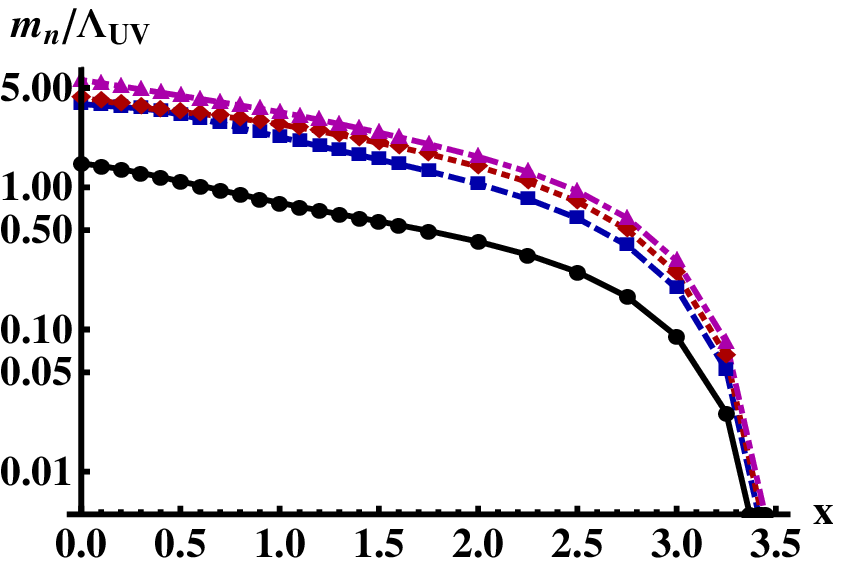}\hfill
\includegraphics[width=0.49\textwidth]{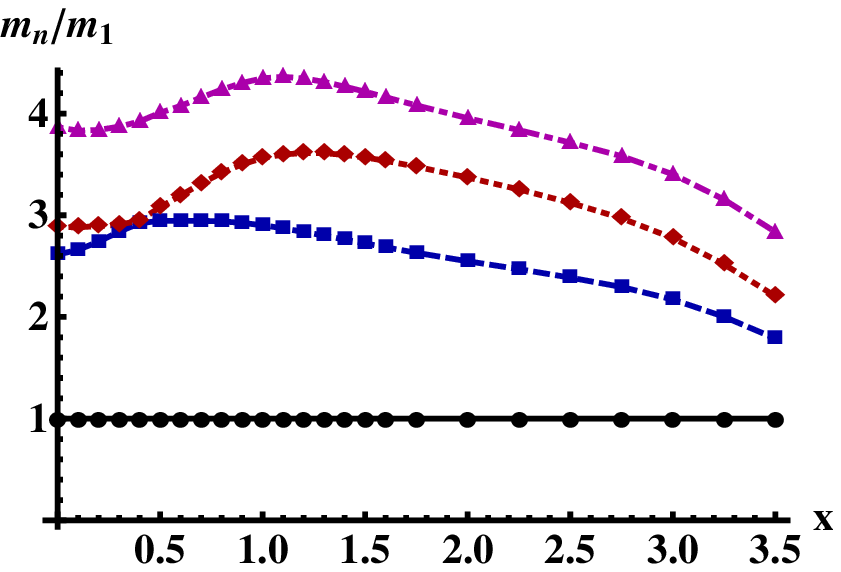}
\end{center}
\caption{\small Singlet scalar meson spectra in the potential II class with SB normalization for $W_0$. They contain the $0^{++}$ glueballs and the singlet $0^{++}$
 mesons that mix here at leading order. Left: the four lowest masses as a function of $x$ in units
of $\Lambda_{\rm UV}$.
Right: the ratios of masses of up to the fourth massive states as a function of $x$.
}
\label{f4}\end{figure}

 \be
{\partial_r \left( V_f(\l,\tau) \gf(\l)^2 e^{A}
G^{-1}\,  \partial_r \psi^V \right)\over {V_f(\l,\tau) \gf(\l)^2 e^{A} G}}
+m_V^2 \psi^V  = 0 \sp G \equiv \sqrt{1+ e^{-2A}\h(\l) (\partial_r \tau)^2 }
\label{vectoreom}
\ee
For the axial vectors we define transverse and longitudinal parts as $A_\m (x^\mu, r)=A^\bot_\m (x^\mu, r)+ A^{\lVert}_\m(x^\mu, r)$
with $A^\bot_\m (x^\mu, r) =  \psi^A(r)\, {\cal A}_\mu (x^\mu)$ and $ \partial^\m {\cal A}_\mu = 0$.
The radial wave function satisfies
\be
{\partial_r \left( V_f(\l,\tau) \gf(\l)^2 e^{ A}
G^{-1}\, \partial_r \psi^A\right)\over V_f(\l,\tau) \gf(\l)^2 e^{ A} G  }
-{4\tau^2 e^{2 A} \h(\l) \over \gf(\l)^2 }\psi^A+m_A^2 \psi^A = 0 \ .
\label{axvectoreom}
\ee
The non-singlet scalar and pseudoscalar fluctuation equations are more complicated and we will present them in \cite{to}.
A few general properties are as follows:

\begin{figure}[!tb]
\begin{center}
\includegraphics[width=0.49\textwidth]{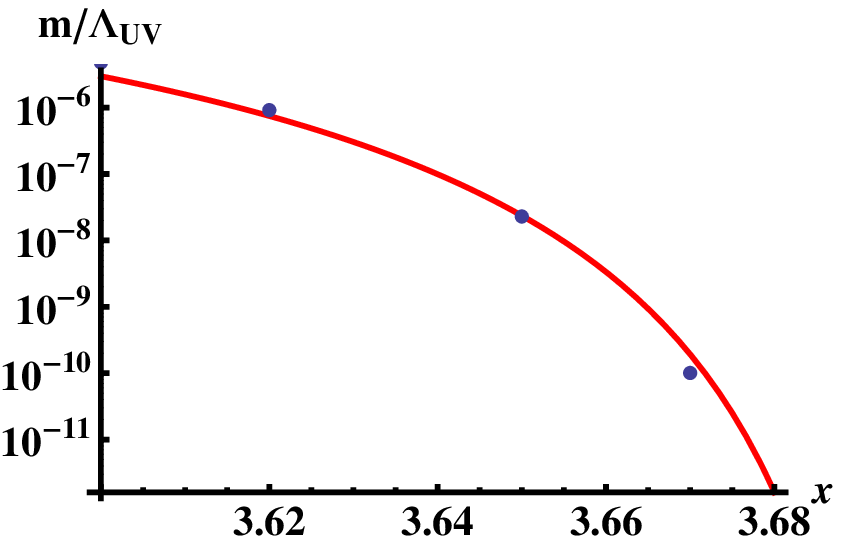}\hfill
\includegraphics[width=0.49\textwidth]{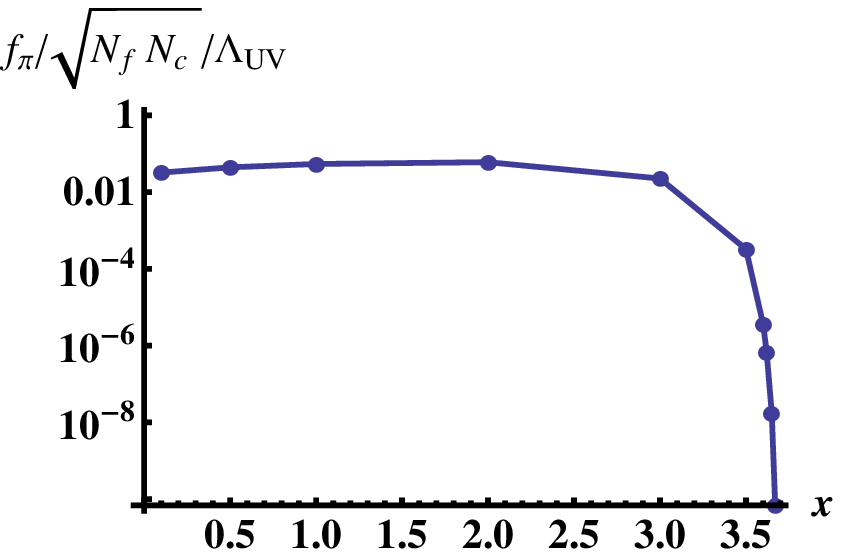}
\end{center}
\caption{\small Potential II with SB normalization for $W_0$. Left: A fit of the $\rho$ mass to the Miransky scaling factor, showing that it displays Miransky scaling in the walking region.
  Right, $f_{\pi}$ as a function of $x$ in units of $\Lambda_{\rm UV}$. It vanishes near $x_c$ following again Miransky scaling.}
\label{f5}\end{figure}

\begin{enumerate}

\item In the conformal window all spectra are continuous.

\item Below the conformal window, $x<x_c$, the spectra are discrete and gapped. The only exception are the $SU(N_f)$ pseudoscalar pions
that are massless, due to chiral symmetry breaking.

\item All masses in the Miransky scaling region (aka ``walking region") are obeying Miransky scaling
$m_n\sim \Lambda_{\rm UV} \exp({-{\kappa \over \sqrt{x_c-x}}})$. This is explicitly seen in the case of the $\rho$ mass in figure
\ref{f5}, left.

\item All non-singlet mass ratios asymptote to non-zero constants as $x\to x_c$.

\end{enumerate}

In figures \ref{f2} and \ref{f3} we present the results for the non-singlet meson spectra (note that the plots on the left of those figures are in logarithmic scale).
The lowest masses of the mesons vary little with $x$ until we reach the walking region. There, Miransky scaling takes over and the lowest masses dip down exponentially fast.
The $\Lambda_\mathrm{UV}$ scale is extracted as usual from the logarithmic running of $\l$ in the UV.

\item The singlet fluctuations, that include the $2^{++}$ glueballs, the $0^{++}$ glueballs and scalar
mesons that mix to leading order in $1/N$ in the Veneziano limit, and the $0^{-+}$ glueballs and the $\eta'$ pseudoscalar tower.
Although the spin-two fluctuation equations are always
simple, summarized by the appropriate Laplacian,
the scalar and pseudoscalar equations are involved
and we refrain from
presenting them here. They will appear in a future publication, \cite{to}.

A few general properties of the singlet spectra are as follows: Both items 1 and 3 above remain as such
(see figure~\ref{f4} for $0^{++}$ scalars).
 Item 2 is replaced by
\begin{itemize}

\item Below the conformal window, $x<x_c$, the singlet spectra are discrete and gapped. The $U(1)_A$ anomaly appears here at leading order and the mixture of the $0^{-+}$ glueball and the $\eta'$ has a mass of ${\cal O}(1)$.

\end{itemize}

We also have:
\begin{itemize}

\item In the scalar sector, for small $x$, where the mixing between glueballs and mesons is small, the lightest state is a meson, the next lightest state is a glueball, the next a meson and so on. However, with increasing $x$, non-trivial mixing sets in and level-crossing seems to be generic.
    This can be seen in figure \ref{f4}, right.

\item All singlet mass ratios asymptote to constants as $x\to x_c$. There seems to be no unusually light state (termed the ``dilaton") that reflects the nearly unbroken scale invariance in the walking region. The reason is a posteriori simple: the nearly broken scale invariance is reflected in the {\em whole} spectrum of bound states scaling exponentially to zero due to Miransky scaling.

\end{itemize}

\end{itemize}

The asymptotics of the spectra at high masses have two different behaviors, depending on the potentials. They are linear (for type  I potentials )
 $m_n^2\simeq c n+\cdots$ or quadratic, $m_n^2\simeq c n^2+\cdots$ for type II potentials.
 There is also the possibility, first seen in \cite{ckp} that the coefficient $c$ in the linear case is different between axial and vector mesons.
 These possibilities do not affect substantially the issues of the dilaton as a well as the S-parameter. 

Finally, let us comment on the possibility of using as background the non-trivial saddle points, found in \cite{jk}, where the tachyon solution has at least one zero (analogous to the Efimov minima).
We have verified explicitly that such saddle points are unstable, as the scalar meson equation has a single mode with a negative mass squared, both in the singlet and non-singlet channels.
This mass is small for small $x$, but becomes large as $x\to x_c$. Therefore the Efimov minima are strongly unstable in the walking regime.

\begin{figure}[!tb]
\begin{center}
\includegraphics[width=0.49\textwidth]{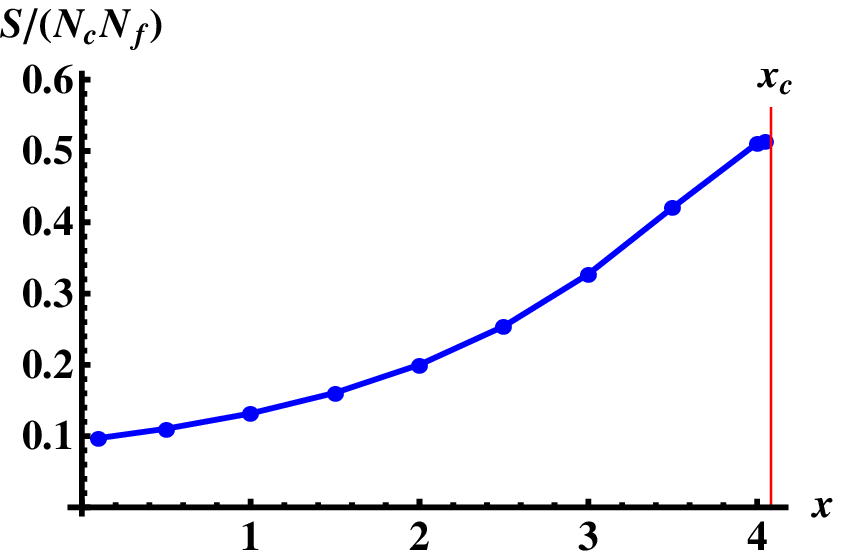}\hfill
\includegraphics[width=0.49\textwidth]{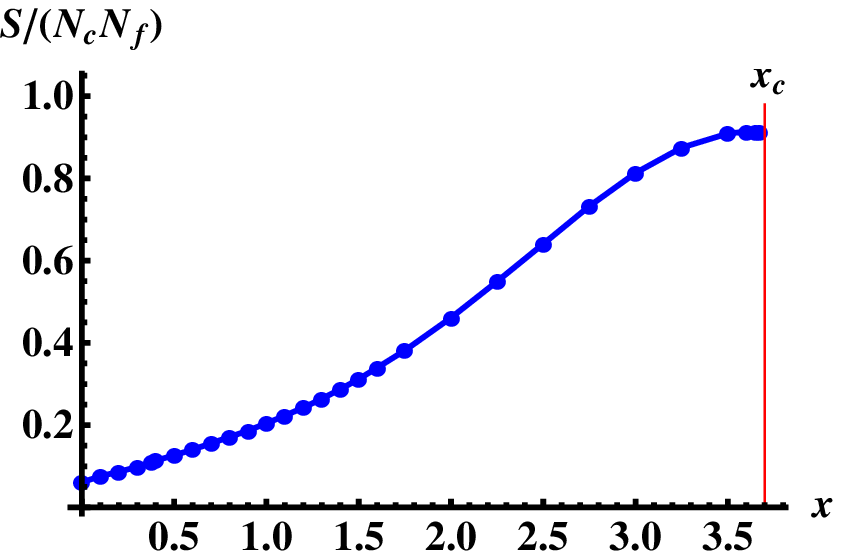}
\end{center}
\caption{\small  Left: The S-parameter as a function of $x$ for potential class I with $W_0={3\over 11}$.
  Right: The S-parameter as a function of $x$ for potential class II with SB normalization for $W_0$.
In both cases $S$ asymptotes to a finite value as $x\to x_c$.
}
\label{f6}
\end{figure}

\section{Two-point functions and the S-parameter}

We have computed the two-point functions of several operators including the axial and vector currents  as well as the scalar mass operator.
We will focus here on the two-point function of the vector and axial currents which can be written in momentum space as
\be
\langle V_{\m}^{a}(q) V_{\n}^{b}(p) \rangle= \Pi^{ab}_{\mu \nu,V,A}(q,p)=-(2 \pi)^4 \delta^4 (p+q)\, \delta^{ab}
\left( q^2\eta_{\m\n}-{q_{\m} q_{\n}}\right)\Pi_{V} (q) \ .
\ee
and similarly for the axial vector. We have  $V_{\m}(x)=\int {d^4 q \over (2\pi)^4} e^{i q x} V_{\m}^{a}(q) t^a \, \psi_V(r)$, where $t^a$, $a=1,\ldots, N_f^2-1$ are the flavor group generators.

Using the expansions
\be
\Pi_A = {f_{\pi}^2 \over q^2} +\sum_{n} {f_n^2 \over q^2 +m_n^2 - i \epsilon}\sp \Pi_V=\sum_{n} {F_n^2 \over q^2 +M_n^2 - i \epsilon}
\label{2}\ee
we determine $f_{\pi}$ as
\be
f_{\pi}^2= -{N_{c} N_{f} \over 12 \pi^2 } \left. {\partial_{r} \psi^A
    \over r} \right|_{r=0,\, q=0} \
\ee
where the normalization was fixed by matching the UV limit of the two point functions to QCD and we chose $\psi^A(r=0)=1$.

A typical example is plotted in figure \ref{f5}, right. The pion scale changes smoothly for most $x$, but is affected directly by Miransky scaling which makes it vanish exponentially in the walking regime.

The S-parameter is
\be
S=4 \pi {d \over dq^2}\left[q^2 (\Pi_V - \Pi_A)\right]_{q=0} =-{N_c N_f \over 3 \pi} {d \over dq^2}\left. \left( {\partial_r \psi^V (r) \over
    r}-{\partial_r \psi^A (r) \over r} \right) \right|_{r=0,\, q=0}
\label{3}\ee
$$
=4\pi\sum_{n}\left({F_n^2\over M_n^2}-{f_n^2\over m_n^2}\right) \ .
$$

As both masses and decay constants in (\ref{2},\ref{3}) are affected similarly by Miransky scaling, the S-parameter is insensitive to it.
Therefore its value cannot be predicted by Miransky scaling alone.
Our results show that generically the S-parameter (in units of $N_fN_c$) remains finite in the QCD regime, $0<x<x_c$ and asymptotes to a finite constant at
$x_c$. The S-parameter is identically zero inside the conformal window (massless quarks) because of unbroken chiral symmetry. This suggests a subtle discontinuity of correlators across the conformal transition.

This behavior of $S$ is in qualitative agreement with recent estimates based on analysis of the BZ limit in field theory \cite{sannino}.
We have also found choices of potentials
where the S-parameter becomes very large as we approach $x_c$.
Our most important result is that generically the S parameter is an increasing function of $x$, reaching it highest value at $x_c$ contrary to previous expectations, \cite{sannino}.

\section{Outlook}

We have analyzed zero temperature spectra of glueballs and mesons in a class of holographic
theories (V-QCD) that are in the universality class of QCD in the Veneziano limit.
 We have verified some generic properties of such spectra in the theory with massless quarks.
 \begin{itemize}

 \item In the conformal window all spectra are continuous.

\item Below the conformal window, $x<x_c$, the spectra are discrete and gapped (except for the pions).

\item All masses in the Miransky scaling region (aka ``walking region") are obeying Miransky scaling
$m_n\sim \Lambda_{\rm UV} \exp({-{\kappa \over \sqrt{x_c-x}}})$.  The same applies to other mass parameters like $f_{\pi}$.

\item All singlet and non-singlet mass ratios asymptote to non-zero constants as $x\to x_c$.
 Therefore there is no  ``dilaton" state. The approximate conformal symmetry is correlated with Miransky scaling instead.

\item For finite values of $x$ there is strong mixing between singlet mesons and glueballs, and occasional level crossings as we vary $x$.

\item The S-parameter in units of $N_fN_c$  is generically ${\cal O}(1)$, is an increasing function of $x$ and asymptotes to a finite constant as $x\to x_c$.
This suggests subtle discontinuities at the conformal transition.

\end{itemize}

The results on the S-parameter suggest that making $S$ arbitrarily small in a walking theory may be more difficult than expected before.
Moreover our results indicate that is probably not the case in QCD in the Veneziano limit.
In \cite{parnachev} that appeared while this work was being finalized, similar conclusions are reached in a different context (probe tachyon-flavor dynamics in
AdS). We find that in the walking region of V-QCD, backreaction of flavor to matter (that is fully implemented here) is important,  among other things,  for the spectra, and therefore the two results are not directly comparable.

 \addcontentsline{toc}{section}{Acknowledgments}
\acknowledgments
This work was in part supported by grants
 PERG07-GA-2010-268246, PIEF-GA-2011-300984, the EU program ``Thales'' ESF/NSRF 2007-2013, and by the European Science
Foundation ``Holograv" (Holographic methods for strongly coupled systems) network.
It has also been co-financed by the European Union (European Social Fund, ESF) and
Greek national funds through the
 Operational Program ``Education and Lifelong Learning'' of the National Strategic
 Reference Framework (NSRF) under
 ``Funding of proposals that have received a positive evaluation in the 3rd and 4th Call of ERC Grant Schemes''.
D.A.
would like to thank the Crete Center for Theoretical Physics for hospitality and the FRont Of
pro-Galician Scientists for unconditional support. I. Iatrakis' work was supported by the project ``HERAKLEITOS II - University of Crete" of the Operational Programme for Education and Lifelong Learning 2007 - 2013 (E.P.E.D.V.M.) of the NSRF (2007 - 2013), which is co-funded by the European Union (European Social Fund) and National Resources.

\end{document}